# Impact of thermo-optical effects in coherently-combined multicore fiber amplifiers

ALBRECHT STEINKOPFF,[1,*] CESAR JAUREGUI,[1] CHRISTOPHER ALESHIRE,[1] ARNO KLENKE,[1,2] AND JENS LIMPERT[1,2,3]

[1] *Institute of Applied Physics, Abbe Center of Photonics, Friedrich-Schiller-Universität Jena, Albert-Einstein-Str. 15, 07745 Jena, Germany*
[2] *Helmholtz-Institute Jena, Fröbelstieg 3, 07743 Jena, Germany*
[3] *Fraunhofer Institute for Applied Optics and Precision Engineering, Albert-Einstein-Str. 7, 07745 Jena, Germany*
*\*albrecht.steinkopff@uni-jena.de*

In this work we analyze the power scaling potential of amplifying multicore fibers (MCFs) used in coherently-combined systems. In particular, in this study we exemplarily consider rod-type MCFs with 2x2 up to 10x10 Ytterbium-doped cores arranged in a squared pattern. We will show that, even though increasing the number of active cores will lead to higher output powers, particular attention has to be paid to arising thermal effects, which potentially degrade the performance of these systems. Additionally, we analyze the influence of the core dimensions on the extractable and combinable output power and pulse energy. This includes a detailed study on the thermal effects that influence the propagating transverse modes and, in turn, the amplification efficiency, the combining efficiency, the onset of nonlinear effect, as well as differences in the optical path lengths between the cores. Considering all these effects under rather extreme conditions, the study predicts that average output powers higher than 10 kW from a single 1 m long Ytterbium-doped MCF are feasible and femtosecond pulses with energies higher than 400 mJ can be extracted and efficiently recombined in a filled-aperture scheme.

## 1. Introduction

Fiber lasers have demonstrated an impressive power performance scaling potential over the last decades, which has opened up a wide field of new applications in the industrial and scientific sectors. However, the onset of mode instability (TMI) [1] and the impact of nonlinearities currently restrict the power scalability of these systems. Large Mode Area (LMA) fibers have enabled further power scaling by distributing the intensity over larger core areas [2]. Although increasing the effective mode area in these fibers has been an effective solution over the last decades, the continued scaling of single-mode core dimensions is technologically challenging. Parallelization followed by beam combination enables scaling beyond single emitter limitations. Numerous techniques have been developed with successful demonstrations employing continuous-wave and pulsed, even ultrafast, lasers [3–8]. While beam-combining has shown its potential to linearly scale the output power with the number of individual amplifying channels used, the potential of this techniques is limited by practicality. Linear scaling of system performance dictates an equivalent requirement for space, number of components and overall complexity. For this reason, high power demonstrations of beam combining have achieved only about one order of magnitude in scaling compared to single emitters [9].

Significant reduction of the footprint, component count and complexity of beam combined systems can be achieved by integrating the individual amplifying channels into a single fiber, a so-called multicore fiber (MCF). Such fibers have already been successfully implemented in optical communications [10], in high temperature sensing systems [11] or in beam shaping systems [12]. Additionally, recent demonstrations have successfully shown that the amplified output of active MCFs can be efficiently coherently combined, with kW-average power [13] and femtosecond-pulse durations [14,15]. These proof-of-principle experiments showed the potential for active MCFs to be used in next generation coherently combined high-power fiber laser systems. In this work, we analyze to which extent the linear power scaling potential of such fibers holds true while paying particular attention to thermal effects that are described in detail in the following sections.

As known, quantum defect heating is one of the most significant heat sources in almost all lasers, so in Yb-doped laser systems. In fiber laser systems in particular, due to the elongated geometry of the fibers, the heat generated in the active cores flows predominantly in the radial direction, which induces a radially inhomogeneous temperature profile as described in the literature [16]. To illustrate the impact of thermal load in an active MCF, the left-hand side of Fig. 1 exemplarily shows a simulated temperature profile at the end facet of a 5x5-MCF operated at high average output power. The underlying simulation tool together with made assumptions will be described in section 2. As it has also been shown in [17], it can be clearly seen that the temperature reaches its maximum at the center of the fiber and, due to heat conduction, decreases towards the outer fiber surface. Due to this radial temperature gradient, which affects the index profile through the thermo-optic effect [16], mode shrinking and mode deformation occur at each core of the MCF, as depicted on the right-hand side of Fig. 1. It can be seen that the modes shrink and are shifted towards the fiber center. These effects depend on the position of each core and can be significantly stronger for the outer cores than for the central ones, which in turn impacts the amplification process as well as the beam combination in a detrimental way.

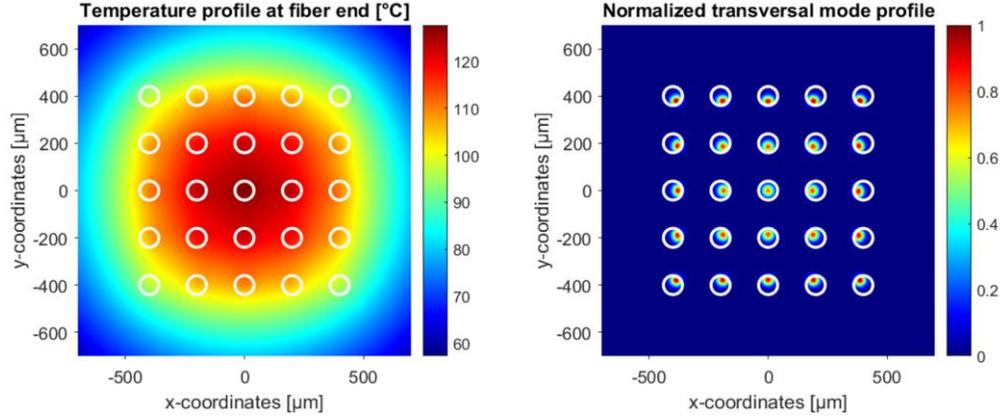

Fig. 1: Exemplary temperature profile (left) and corresponding transversal intensity profile (right) in a 5x5 active MCF under heat load, with cores highlighted as white circles.

Worth to be mentioned, the presented analysis is, to a large extent, valid for arbitrarily arranged cores and beam combination in general. However, we concentrate on the mentioned squared pattern of cores and filled aperture coherent combination as described and experimentally demonstrated in [18]. Coherent combining is a phase-stabilized superposition of the E-fields, which leads to a summation of the power emitted by the individual emitters, in the case of a MCF, the individual cores. However, as shown later in this letter, the thermally induced non-uniform mode shrinking between the different cores (see Fig. 1 right) will affect the spatial overlap, with it the combining efficiency and finally the combined output power. Also, the output power emitted by the individual cores might vary, since the amplification efficiency depends on the spatial overlap of the modes with the doped cores. Moreover, the thermally induced modifications of the different cores will result in different optical path lengths and a different accumulated nonlinear phase, which could be important for the coherent combination in ultrashort pulse laser systems. In this work we will pay particular attention to these effects and, especially, we will evaluate the maximum extractable and combinable output power/energy. In the first section, the simulation tool together with the fiber parameters will be described. Subsequently, the different effects originating from the non-uniform temperature profile will be explained. In the last section, the simulation results will be presented and discussed together with the power and pulse energy scaling prospects.

## 2. Simulation tool and fiber parameters

A simulation tool has been developed in order to describe an MCF amplifier system accounting for all the previously mentioned effects. The working principle of this tool is schematically depicted in Fig. 2. It iteratively solves the laser rate equations, from which the heat-load can be calculated, whereby we assume that it is caused exclusively by the quantum defect. This results in a temperature profile that leads to a modification of the refractive index profile and, with it, to the distortion of the guided core modes along the fiber. Since single transverse mode operation in the individual cores of the MCFs is assumed, only the fundamental modes (FM) in each of the cores are taken into account. With the assumption that no optical coupling between the cores occurs (due to the large core-to-core distance, see [19]) the propagating modes can be calculated by solving the scalar wave equation for each core, as described in [20], at every point along the fiber. The distortion of the propagating modes along the fiber influences, in turn, the solution of the rate equations, which must then be recalculated. This cycle is repeated until convergence is reached. As convergence criterion, a maximum deviation of 1 % in the temperature, signal and pump power evolution between two iteration steps has been chosen. The result from the simulation contains the 3-dimensional temperature and refractive index profiles, the power evolution of the signal light in each core, the pump power evolution as well as the electrical field distribution in each core along the fiber.

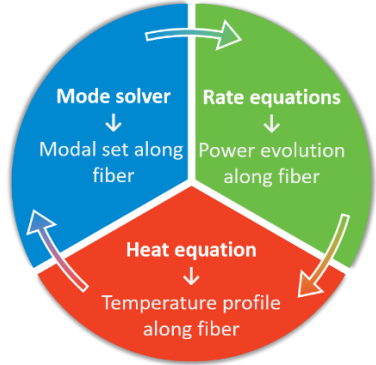

Fig. 2: Schematic flow-diagram of the simulation tool.

The simulation tool is capable of solving fiber structures with arbitrary core dimensions, core arrangements and index profiles. In this work we consider configurations of MCFs with a quadratic arrangement of step-index cores (as shown in Fig. 3), from 2x2 up to 10x10 cores. The square pattern is compatible with the typical splitting and combining elements presented in [18], enabling an efficient coherent combination of the emitting cores. The simulated ytterbium-doped MCFs are counter-pumped, whereas the performance of the fibers at different pump power levels is analyzed.

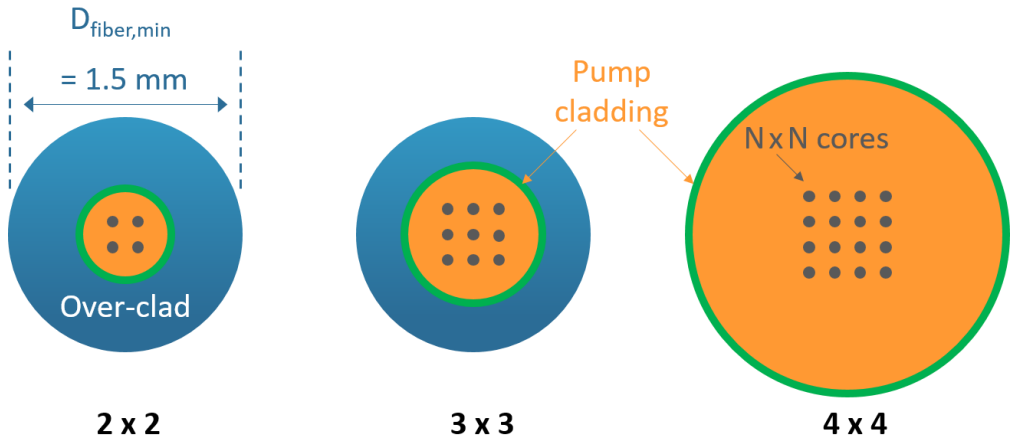

Fig. 3: Schematics of MCFs with different core and cladding configurations. The pump cladding (orange) with a low index layer (green) is adapted in each simulated case to achieve the same small signal pump absorption of 20 dB/m. The outer fiber diameter is at least 1.5 mm, but when the pump cladding (orange) is larger than this value, the outer fiber diameter has the same dimension as the pump cladding.

The fiber length has been chosen to be 1 m. This is a typical length of rod-type fibers which are suitable for high-energy, ultrafast pulse operation. The diameter of the active cores (with an Yb-doping concentration of $5 \cdot 10^{25}$ ions/m$^3$) in the different designs are 30, 50 and 80 µm respectively. All cores possess a flat step-index profile (in cold operation) that corresponds to a V-parameter of 3. The pump cladding consists of silica glass and its size is appropriately chosen to achieve a small-signal pump absorption of 20 dB/m at 976 nm wavelength in all cases. The green ring indicates a low index layer to obtain pump guidance within the orange area (pump cladding). The fiber diameter has at least the same dimension as the cladding size. However, in the case of the smaller pump claddings (i.e those with a diameter smaller than 1.5 mm), an additional silica over-clad is added to obtain a mechanically stable rod-type fiber of at least 1.5 mm outer diameter (blue ring in Fig. 3). Furthermore, we would like to point out that, as usual for rod-type fibers, no additional polymer coating is considered in these simulations. The coolant medium at the fiber outer surface is chosen to be water (with a thermal conductivity of 0.58 W/(m·K) and a temperature of 25 °C).

## 3. Thermal effects and their consequences

As mentioned before, the typical non-uniform thermal transverse profile in an MCF, shown in Fig. 1 (left), will lead to different propagation constants along the individual cores and, thus, to different optical paths lengths. Our simulations reveal that the maximum optical path difference among all herein considered scenarios, encountered in a 1m long 10x10 80 µm MCF with 300 W of average output power being extracted per core, is smaller than 25 µm, which is low enough to be compensated by phase stabilization systems, as done in [18].

A more serious impact of the thermal profile, shown in Fig. 1 (right), is mode shrinking, which also varies from core to core. An example of the evolution of the effective mode area of the propagating fundamental modes (FM) in the different cores (color-coded) is depicted in Fig. 4 for a 5x5 MCF, with 80 µm cores, counter pumped at 976 nm wavelength with an average output power of 3.4 kW.

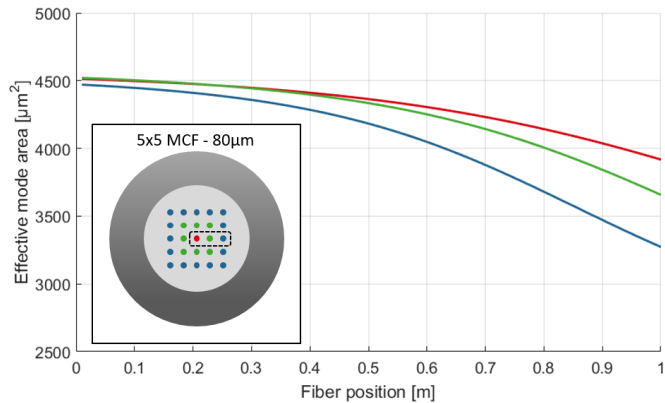

Fig. 4: Evolution of the effective mode area of the propagating fundamental modes in three different cores (indicated by the dashed rectangle in the inset) along a 5x5 MCF with 80 µm cores, counter pumped with an average output power of 3.4 kW. The inner core (red line – also colored red in the inset) shows less shrinking than the middle core (green) and the outer core (blue).

It can be seen that the propagating FM in the inner cores of the MCF (red and green curve) show significantly less mode shrinking than the one in the outer core (blue curve). In this example a maximum difference of ~20 % in the mode area at the fiber end facet between the inner and the outer core is obtained. The different distribution of the intensity profiles in the cores at the fiber end facet will influence the spatial combining efficiency $\eta_{comb,spatial}$ that is calculated as follows:

$$\eta_{comb,spatial} = \frac{\int \left| \sum_i^{N^2} \sqrt{P_i} \cdot E_{i,norm}(x-x_i, y-y_i) \right|^2 \cdot dxdy}{N^2 \cdot \sum_i^{N^2} \int \left| \sqrt{P_i} \cdot E_{i,norm}(x-x_i, y-y_i) \right|^2 \cdot dxdy} \quad (1)$$

$\eta_{comb,spatial}$ can be understood as the ratio of the combined power to the sum of the power emitted by all $NxN$ cores. The combined power is calculated by spatially superimposing the normalized electrical field amplitudes $E_{i,norm}$ emitted by the individual $NxN$ cores with the core centers $x_i$ and $y_i$ at the fiber end facet (which are part of the solution given by the simulation tool). At this point it is important to note that different output powers $P_i$ emitted by each core might occur. This is due to the fact that the different mode sizes along the fiber will, in principle, influence the amplification behavior in each core (due to the different overlap with the doped core area). This will also be considered in the calculation of the combining efficiency.

Fig. 5 shows the intensity profiles of a 5x5 MCF with 80 μm cores at different power levels. On the left side (corresponding to 500 W of total output power) almost no mode shrinking occurs, which results in a high spatial combining efficiency ($\eta_{comb,spatial}$) of nearly 100 %. However, at high output power levels (3400 W in total, as shown on the right-hand side) significant mode shrinking occurs. Since this mode deformation is not homogeneous for all cores, the combining efficiency drops to ~75 % in this case.

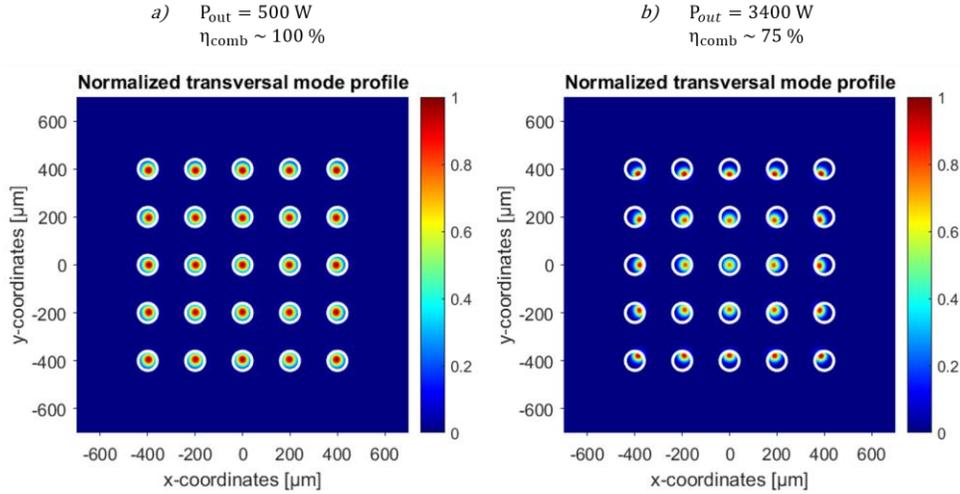

Fig. 5: Example of intensity profiles in a 5x5 MCF (80 μm cores) at different power levels. In a) ~100% of combining efficiency is achieved whereas in b) due to non-uniform mode shrinking just ~75% of combining efficiency is obtained.

Another important parameter that has to be taken into account, is the accumulated nonlinear phase for each core. This parameter can play a significant role in coherently combined ultrashort laser systems (where high peak powers occur, even with stretched pulses), since it might lead to different temporal phases between the single emitters. As described in [21], the so-called B-integral for the $i^{th}$ core can be calculated as follows:

$$B_i = \frac{2\pi}{\lambda} \int_0^L n_2 I_{peak,i}(z) dz \qquad (2)$$

With $\lambda$ being the signal light wavelength, $n_2$ the nonlinear index and $I_{peak,i}(z)$ the peak intensity of the propagating pulse along the core *i*. Here, the strength of the simulation tool can be seen, since the mode shrinking along the fiber, and with it the change in peak intensity, are taken into account. As shown in Fig. 4, the mode shrinking is different in each core, leading to different B-integrals, which are also be accounted for in the considered coherent combination. A detailed analysis on the impact of B-integral variations among the interference partners on the combining efficiency is given in [22].

## 4. Results and discussion

As described in the last section, there are various effects that must be considered when operating an MCF amplifier at high thermal loads. First, prohibitively high temperatures might occur due to the extremely high powers that can be extracted from these waveguides. In Yb-doped fibers, in particular, the optical properties (such as the absorption and emission cross sections, as described in [23]) will drastically change when the fiber exceeds a temperature of 500 °C. Especially the pump absorption from 900 to 1000 nm and the emission above 1010 nm decreases while the re-absorption in the signal band increases. Moreover, thermally induced damage can start to happen at around 1000 °C - as additionally described in [24]. Following these studies, a maximum occurring temperature of 500 °C in the MCF will be considered as an upper temperature limit of the safe working range in our simulations. Another thermal effect that must be taken into account is transverse mode instability (TMI) that occurs at high average output powers (in each core). Previous experiments with single- and multicore fibers have shown that the TMI threshold for ~1 m long fibers can be estimated to occur at ~300 W per core [25,26], while the threshold of each single core is not changed by the number of cores [19]. Moreover, it has been theoretically shown that the TMI threshold does not significantly change with the core size when the V-parameter is kept constant [27,28]. Therefore, in our simulations we define a maximum extractable average power per core of 300 W for all core sizes as another performance limit due to TMI.

Taking all these limitations into account leads to the results as shown in Fig. 6, where the maximum occurring temperature in the fiber is plotted as a function of the extracted power per core for different MCF designs ranging from 2x2 to 10x10 cores (30 µm core diameter on the left and 80 µm core diameter on the right-hand side). It turns out that the maximum temperature always occurs in one of the cores since the heat (due to the quantum defect) is generated there. It can be seen that this maximum temperature rises almost linearly with the extracted power per core in all cases. For any fixed extracted power per core, the absolute temperature also rises with an increasing number of cores (from 2x2 up to 10x10) since much more total power is extracted from the same fiber length. In most cases the stable operation regime of the fiber (highlighted as a green area in Fig. 6) is limited by the TMI threshold when the extracted power per core reaches 300 W. Consequently, this corresponds to a total average output power (taking all cores into account) ranging from 1.2 kW for the 2x2 MCF to almost 30 kW for the 10x10 MCF - in the case of 80 µm cores. However, in a few cases - especially for MCFs with a smaller core diameter of 30 µm and high number of cores - it can be seen that the thermal limit threshold can be reached before the TMI limit. Due to their smaller fiber and core dimension, MCFs with 30 µm cores reach a higher maximum temperature at the same extracted power compared to 80 µm cores. Regarding a simple step-index-fiber this behavior was analytically shown and explained in [16] and is analogous for MCFs. Thus just 220 W, 200 W, 170 W and 150 W of average output power per core (instead of 300 W) can be reached with the 7x7, 8x8, 9x9 and 10x10 MCF, respectively, when considering 30 µm core diameter.

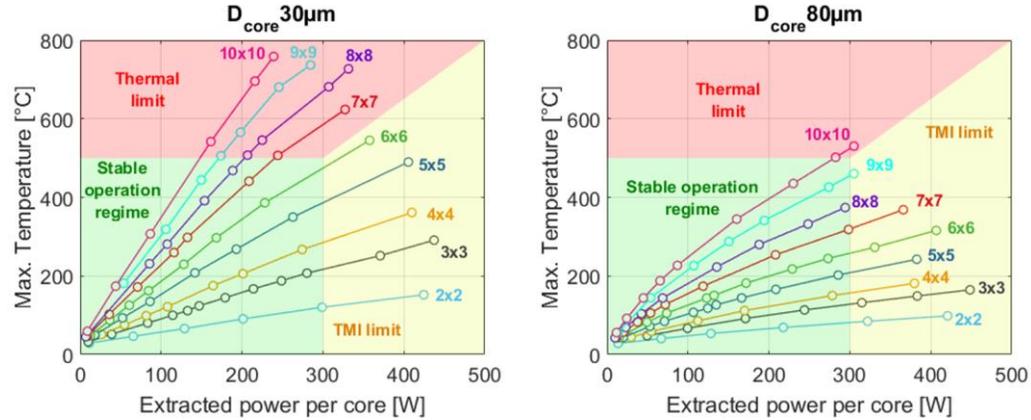

Fig. 6: Max. temperature occurring in the Yb-doped MCFs ranging from 2x2 up to 10x10 cores vs. extracted power per core with (left) 30 µm and (right) 80 µm core diameter. The green, red and yellow areas represent the regimes of stable operation, thermal limit and onset of TMI, respectively.

Another important aspect beside the maximum extractable output power per core is the combining efficiency, as described in the previous section. This parameter has to be taken into account for coherently combined systems since it limits the combinable output power emitted by an MCF. To stay in an acceptable working regime we additionally define a self-imposed performance limit of a minimum combining efficiency of 70 %.

The combining efficiency as a function of the extracted power per core – considering MCFs with 30 µm (left) and 80 µm (right) core diameter - is shown in Fig. 7. It can be seen that the combining efficiency decreases with more extracted power per core for both core diameters. This is due to the non-uniform mode shrinking/deformation shown in Fig. 5. Moreover, the degradation of the combining efficiency is worse for larger cores since their propagating modes (for a constant V-parameter) are more sensitive to thermally induced changes. Additionally, it can be seen that using more cores per fiber will also lead to a decrease in the combining efficiency at the same extracted power per core. This is because more cores generate more total heat in the fiber, which will lead to a higher distortion of the modes at the fiber end.

It can be seen that, e.g. with a 6x6 MCF with 30 µm cores a combining efficiency of ~ 93 % is reached at an extracted power of 300 W per core, which corresponds to a total extracted and combined output power of ~ 10 kW.

When combining the results from Fig. 7 with those of Fig. 6, it can be seen that average power scaling of MCFs with smaller core dimensions (and high number of cores) is limited by the maximum occurring temperature, whereas MCFs with large core dimensions are instead restricted by the lower limit of the combining efficiency of 70 % (see Fig. 7 right).

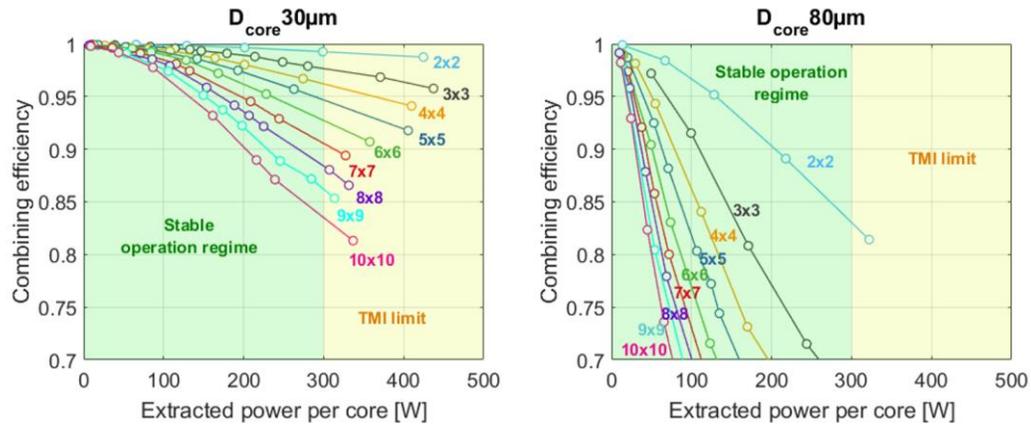

Fig. 7: Combining efficiency vs. extracted power per core for different MCFs with (left) 30µm and (right) 80µm core diameter. The green and yellow shaded areas represent the regime of stable operation and TMI, respectively.

Taking all the limitations simultaneously into account (i.e. the maximum occurring temperature of 500 °C, the maximum extractable output power per core of 300 W due to TMI and a minimum combining efficiency of 70 %) will lead to a maximum combinable output power for all the different MCF configurations with different core sizes, as shown in Fig. 8. Thus, the maximum combinable output power for 30, 50 and 80 µm MCFs is depicted by the red, green and blue curves, respectively. The red shaded area shows the operating regime above our theoretical maximum (i.e. considering an extracted output power per core higher than 300 W together with a combining efficiency of 100 %). With more cores, the maximum combinable output power starts rising linearly for the different core sizes (from 30 µm to 80 µm). However, large core MCFs show a worse performance in terms of the combinable output power since the combining efficiency decreases drastically, as shown in Fig. 7 (right). Additionally, it can be seen that the linear rise of the combinable output power for the 30 µm MCF flattens for 7x7 cores and more. At this point a maximum temperature of 500 °C is observed in the fiber before the maximum extracted power per core of 300 W is reached, as depicted in Fig. 6 (left). The same flattening-effect can be observed with the 50 µm MCF with more than 9x9 cores which leads to the same performance in terms of average power as the 30 µm MCFs.

In spite of this, it can be seen that more than 10 kW (with at least a 6x6 30 µm MCF) of combinable output power can be realized with such a 1m long MCF arrangement. Please note that this presented study is rather an extreme scenario and that the combinable average power can be further scaled by using longer fibers since the thermal load is distributed over a longer fiber length.

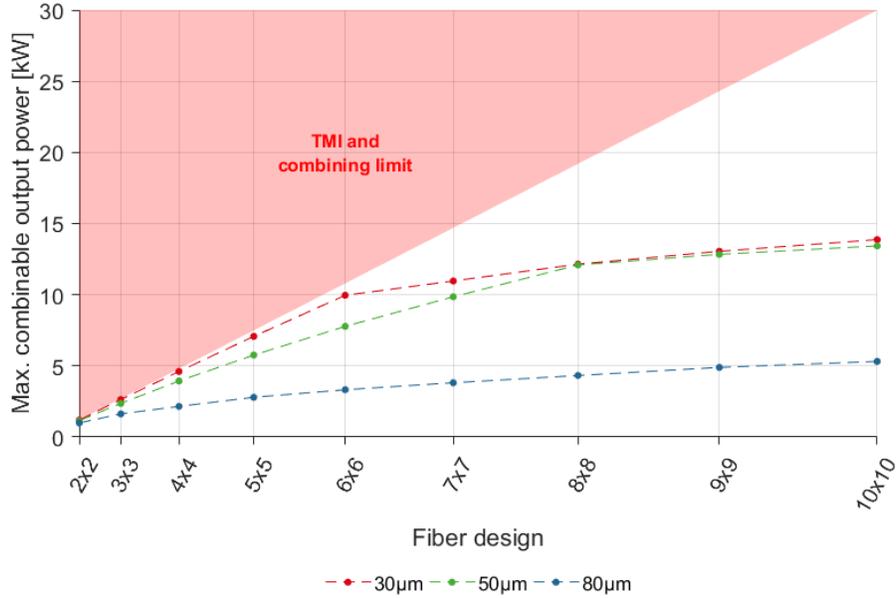

Fig. 8: Maximum combinable output power for the different MCFs (2x2 to 10x10) with different core sizes (30, 50 and 80 µm). The red shaded area represents operation beyond the TMI threshold of 300 W per core together with a 100 % combining efficiency.

The results shown above are valid for CW-operation in MCF systems. With pulsed operation (and, especially, ultrafast pulse operation) the accumulated nonlinear phase - described in equation (2) - must be accounted for since it will impose a further performance limit. Thus, instead of using the CW-power along the fiber the evolution of the pulse peak power must be considered. In order to do this, it will be assumed that the peak power of the pulses follows the same amplification profile along the fiber as the CW power calculated by the simulation tool. In this case Gaussian pulses with a stretched duration of 10 ns are assumed (this stretched pulse duration, although large, is something that will be implemented in the next generation of state-of-the-art, high-performance ultrafast fiber laser systems). This information, together with the repetition rate (that will be varied in this study) allows the calculation of a corresponding pulse energy and B-integral.

The nonlinear accumulated phase will in most cases distort the temporal profile of the recompressed pulses. Using a spectral shaping device is a well-known technique to reduce the impact of the temporal Kerr nonlinearity, the B-integral [29,30]. Even though this approach has achieved good results in the past, compensating for high B-integrals is technologically challenging [5,31]. Hence, we chose a maximum B-integral of 10 rad as the performance limit in the following consideration.

Taking this limitation into account (in addition to the others already considered in the CW study) will lead to a maximum combinable and extractable pulse energy for the different MCFs at different repetition rates, as shown in Fig. 9. Hereby it can be seen that, with decreasing repetition rate (at the same combined output power), the pulse energy rises since it is only limited by the maximum combinable output power, as seen in Fig. 8. However, at a certain point a maximum B-integral of 10 rad or more (red shaded area) is reached which limits any further scaling of the pulse energy. It can be seen that for 30 µm MCFs (left) the maximum B-integral is reached already at 100 kHz whereas 80 µm MCFs (right) allow for repetition rates as low as 10 kHz. The larger mode area in 80 µm cores consequently leads to significantly higher pulse energies. At this point it is important to note that the different B-integrals for all cores have been taken into account to calculate the impact on the combining efficiency, as demonstrated in [22]. It turns out that, even at the highest pulse energies, the combining efficiency is just affected by a very few percent and, therefore, the impact of the B-integral on this parameter is neglected in the presented considerations.

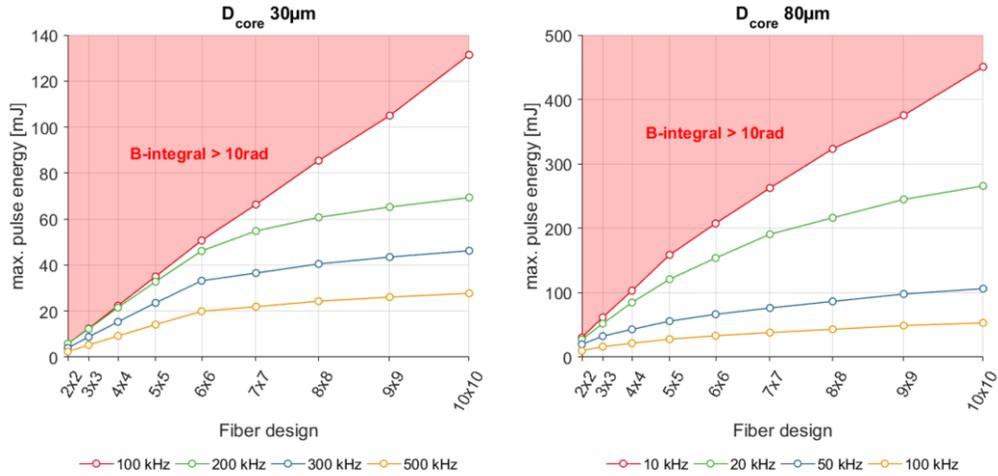

Fig. 9: Maximum achievable pulse energy at different repetition rates for 30µm (left) and 80µm (right) MCFs.

The maximum extractable and combinable pulse energy for MCFs with different core dimensions (30, 50, 80 µm) at their optimum repetition rate (i.e. that corresponding to a maximum occurring B-integral of 10 rad, as shown in Fig. 9) is depicted in Fig. 10. It can be seen that using a 10x10 MCF with 30 µm core dimensions (left figure) will allow for a maximum pulse energy of ~ 130 mJ at 100 kHz repetition rate. On the other hand, using a 10x10 MCF with 80 µm core dimensions will result in a maximum extractable and combinable pulse energy of ~ 450 mJ when using a repetition rate of 10 kHz.

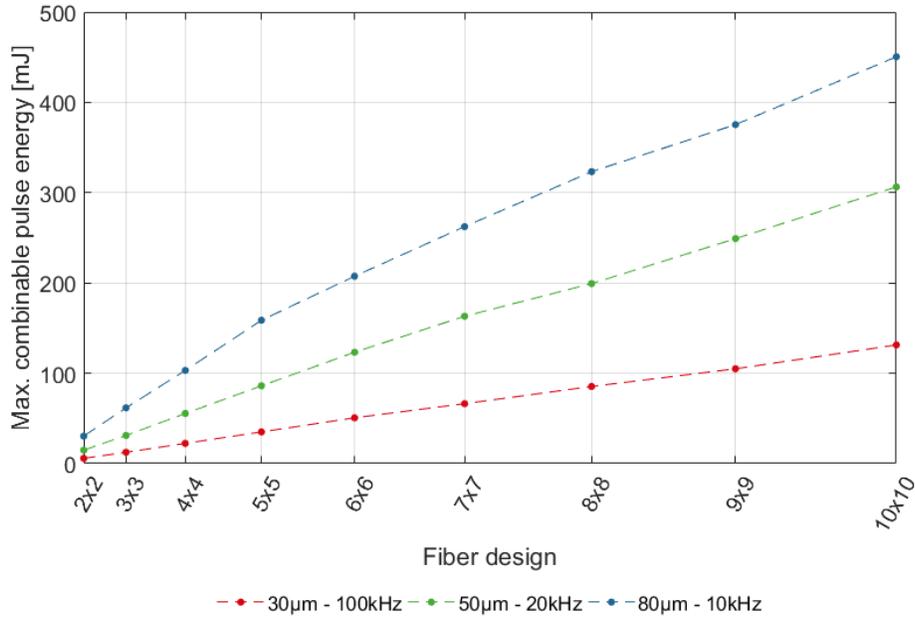

Fig. 10: Maximum achievable pulse energy (at the optimum repetition rate) for all simulated MCFs with different core dimensions.

## 5. Conclusion

Simulation results for coherently combined ytterbium-doped multicore fiber amplifier systems have been presented. In order to do this, a tool has been developed that numerically solves and analyzes the amplification behavior in MCFs together with thermal considerations. Using this model allows predicting the expected thermal load and temperature-dependent effects such as mode-shrinking as a function of MCF design parameters. It turns out that the combining efficiency is particularly strongly influenced by the thermally-induced mode shrinking and deformation that differs from core to core – depending on its position.

Due to the limitations imposed by thermal effects, the best strategy for scaling the average power appears to be increasing the number of cores and using relatively small core diameters. By doing so, our simulations suggest that it should be possible to extract and combine around 13.5 kW from a just 1 m long 10x10 MCF with 30 µm core dimensions. The use of larger core diameters, on the other hand, will help to maximize the extractable and combinable pulse energy in these systems in ultrashort pulse operation. Assuming a stretched pulse duration of 10 ns and a maximum B-integral of 10 rad, it was possible to show that femtosecond pulses with up to 450 mJ of pulse energy in a 10x10 MCF with 80 µm core dimensions together with an average power of 4.5 kW can be achieved. At this point we want to emphasize that our study does not account for other effects that might reduce the combining efficiency in MCFs, such as multimode operation, tolerances in core sizes/positions or polarization changes. According to already existing coherently combined systems, an additional reduction of power/energy performance of 10-20 % (due to a plethora of spurious effects, such as multimode operation of the cores, imperfections in the optical elements, etc.) has to be taken additionally into account for a realistic estimation. According to our simulations, considering the aforementioned power and energy scaling prospects together with these additional limitations, a combined average power of more than 10 kW in CW operation and more than 400 mJ (4.0 kW) in pulsed operation seem feasible with MCF systems.

Future related studies will focus on experimental investigations of the thermal effects that have been revealed and presented in this work. Moreover, strategies have to be worked out to mitigate the impact of thermal effects in MCFs so that the combined average output power and pulse energy can be further increased. This will be an important step towards compact ultrashort pulse Joule-class MCF laser systems in the future.

**Acknowledgment**

The research leading to these results has received funding from the European Research Council under the ERC grant agreement no. [670557] 'MIMAS', no. [835306] 'SALT', the Free State of Thuringia, the European Social Fund, the German Federal Ministry of Education and Research (funding program Photonics Research Germany, contract no. 13N15244), the Deutsche Forschungsgemeinschaft (DFG, German Research Foundation) – 416342637 and the Fraunhofer Cluster of Excellence 'Advanced Photon Sources'.

**Disclosures**

The authors declare no conflicts of interest.